\documentclass[11pt,a4paper]{article}
\usepackage{jcappub} \notoc
\usepackage{amsmath}
\usepackage{amsfonts}
\usepackage{multirow}

\newcommand{\Neff}{N_\text{eff}}

\begin{document}


\title{
A new life for sterile neutrinos:\\
{\huge \mbox{resolving inconsistencies using hot dark matter}\footnote{Based on observations obtained with Planck (\texttt{http://www.esa.int/Planck}), an ESA science mission with instruments and contributions directly funded by ESA Member States, NASA, and Canada.}}\\
}

\author[a]{Jan Hamann}
\author[b,c]{Jasper Hasenkamp}

\affiliation[a]{Theory Division, Physics Department\\
 CERN, CH-1211 Geneva 23, Switzerland}

\affiliation[b]{Center for Cosmology and Particle Physics, Physics Department\\
 New York University, New York, NY 10003, USA}

\affiliation[c]{II.~Institute for Theoretical Physics\\
 University of Hamburg, 22761 Hamburg, Germany}

\emailAdd{jan.hamann@cern.ch, jasper.hasenkamp@nyu.edu}


\abstract{
Within the standard $\Lambda$CDM model of cosmology, the recent Planck measurements have shown discrepancies with other observations, e.g., measurements of the current expansion rate $H_0$, the galaxy shear power spectrum and counts of galaxy clusters.
We show that if $\Lambda$CDM is extended by a hot dark matter component, which could be interpreted as a sterile neutrino, the data sets can be combined consistently.
A combination of Planck data, WMAP-9 polarisation data, measurements of the BAO scale, the HST measurement of $H_0$, Planck galaxy cluster counts and galaxy shear data from the CFHTLens survey
yields $\Delta N_\mathrm{eff} = 0.61 \pm 0.30$ and $m_\mathrm{s}^\mathrm{eff} =  (0.41 \pm 0.13)~\mathrm{eV}$ at $1\sigma$.
The former is driven mainly by the large $H_0$ of the HST measurement, while the latter is driven by cluster data. CFHTLens galaxy shear data prefer $\Delta \Neff >0$ and a non-zero mass.
Taken together, we find hints for the presence of a hot dark matter component at $3\sigma$.  A sterile neutrino motivated by the reactor and gallium anomalies appears rejected at even higher significance and an accelerator anomaly sterile neutrino is found in tension at $2\sigma$.
}

\maketitle

\section{Introduction \label{sec:intro}}

The measurements of the cosmic microwave background (CMB) temperature anisotropies by the Planck satellite have set a new standard of precision in cosmology.
One of the central results of the Planck data is that CMB data by themselves appear to be perfectly well described by the standard $\Lambda$CDM {\it vanilla} model,
and show no preference for extended models~\cite{Ade:2013zuv}. 
However, it has also been noted that within the vanilla model, several other cosmological observations appear to be inconsistent with CMB data at the 2-3$\sigma$ level.
Notably, the expansion rate $H_0$ found from measurements of type Ia supernov\ae\ and Cepheid variables~\cite{Riess:2011yx} is larger than the
one preferred by CMB data.  Also, the amount of power in the matter perturbation at small scales, characterised by the parameter $\sigma_8$, inferred from CMB data is
larger than the $\sigma_8$ deduced from the Planck galaxy cluster count~\cite{Ade:2013lmv} and measurements of the shear power spectrum by the CFHTLens
collaboration~\cite{Heymans:2013fya}.

While these inconsistencies might indicate unresolved systematic effects in the data due to an incomplete understanding of astrophysics, one should keep
in mind that any ``inconsistency between measurements'' is a model-dependent statement as long as different quantities are measured.  
Therefore, these inconsistencies might also indicate a problem with the vanilla model and turn out to be hints at new particle physics.
In Ref.~\cite{Ade:2013zuv}, the former interpretation was chosen, and the discrepant data sets were consequently not combined with CMB data.  In this work
we shall subscribe to the latter interpretation, and consider an extension of the vanilla model by a hot dark matter (HDM) component, also known as $\Lambda$
Mixed Dark Matter ($\Lambda$MDM) model.  The HDM component is characterised by two new parameters: its total energy density and the mass of the HDM particle.

Our motivation for this particular model is twofold:  firstly, this model is a good candidate for resolving the aforementioned data inconsistencies.
At decoupling, the HDM component is relativistic and hence contributes to the radiation energy density, parameterised by the effective number of neutrino species, $N_\mathrm{eff}$.
Since $\Neff$ is positively correlated with $H_0$~\cite{Bashinsky:2003tk,Hou:2011ec,Julienbook}, an increase in $\Neff$ should ameliorate the tension with the larger values found in local observations.
At later times, during structure formation, the presence of HDM will inhibit the growth of structures below its free-streaming scale, thus reducing power at small scales and leading to
lower values of $\sigma_8$, which could improve consistency with cluster count and galaxy shear data (see also Ref.~\cite{Burenin:2013wg} for a pre-Planck discussion).
Secondly, the $\Lambda$MDM model also encompasses the light sterile neutrino scenario, which has been suggested to resolve the 
accelerator~\cite{Athanassopoulos:1996jb,Aguilar:2001ty}, reactor~\cite{Mention:2011rk} and gallium~\cite{Abdurashitov:2005tb,Giunti:2012tn}
anomalies in neutrino oscillation experiments. 
The reactor and gallium anomalies jointly prefer new mass-squared differences $\Delta m^2 \gtrsim 1~\text{eV}^2$~\cite{Abazajian:2012ys}, while the various accelerator experiments~\cite{Armbruster:2002mp,Aguilar-Arevalo:2013pmq,Antonello:2012pq} prefer $\Delta m^2 \sim 0.5~\text{eV}^2$. In all three cases mixing angles are preferred that generically lead to an increase of $\Delta \Neff=1$.

The paper is organised as follows: in Sec.~\ref{sec:modelanddata} we briefly present the cosmological models and data that we use to obtain the results provided and discussed in Sec.~\ref{sec:results}. 
We conclude in section~\ref{sec:conclusions}.

\section{Models and Data \label{sec:modelanddata}}
\subsection{Models}
We consider two different cosmological models: the \textit{vanilla} $\Lambda$CDM ($\Lambda$: dark energy and CDM: cold dark matter) and an extension of that model by an additional hot dark matter (HDM) component. This extended model is also known as  $\Lambda$MDM as the dark matter consists of a cold (C) and a hot (H) component or, in other words, it is a mixture (M) of the two components  in addition to the ordinary neutrinos.  Since we parameterise the HDM itself by cosmological sterile neutrino parameters, we will also refer to the extended model shortened also as ``sterile model".

For these models we follow the parameterisations chosen in Ref.~\cite{Ade:2013zuv}. The free parameters (and their respective prior distributions) are listed in Table~\ref{tab:parameters}.  In addition, we vary 13 nuisance parameters required for modelling the Planck data, as described in Ref.~\cite{Planck:2013kta}.  From these base parameters, we can derive a number of other interesting parameters, e.g., the current expansion rate of the Universe $H_0$, the root-mean-square matter fluctuations in $8\, h^{-1}$~Mpc spheres today computed in linear theory, $\sigma_8$, and the current matter energy density in units of the critical energy density $\Omega_\text{m} = \Omega_\text{cdm} + \Omega_\text{hdm} + \Omega_\mathrm{b} = (\omega_\text{cdm} + \omega_\text{hdm} + \omega_\mathrm{b})/h^2$.
We adopt the usual convention of writing today's Hubble parameter as $H_0 = 100 \, h \text{ km } \text{s}^{-1} \text{ Mpc}^{-1}$ and fix the sum of neutrino masses $\Sigma m_\nu = 0.06$ eV to take the minimal value indicated by global fits to recent neutrino oscillation and other data~\cite{Tortola:2012te}.  The Big Bang Nucleosynthesis consistency relation~\cite{Hamann:2007sb} is imposed to fix the primordial Helium fraction.

\begin{table*}[t]
  \caption{Physical parameters and prior ranges of the models considered. \label{tab:parameters}}
\begin{center}
{\footnotesize
  \hspace*{0.0cm}\begin{tabular}
  {c|cc} \hline \hline
  Parameter 						& Symbol 								& Prior range		\\	\hline	
  Baryon density					& $\omega_\mathrm{b}$					& $[0.005,0.1]$		\\
  Cold dark matter density				& $\omega_\mathrm{cdm}$ 				& $[0.001,0.99]$	\\
  Sound horizon parameter			& $\theta_\mathrm{MC}$					& $[0.5,10]$		\\
  Reionisation optical depth			& $\tau$								& $[0.01,0.8]$		\\
  Scalar spectrum amplitude			& $\log \left[ 10^{10} A_\mathrm{s} \right]$		& $[2.7,4]$		\\
  Scalar spectral index				& $n_\mathrm{s}$						& $[0.9,1.1]$		\\ \hline
  Extra radiation degrees of freedom		& $\Delta N_\mathrm{eff}$				& $[0,2]$			\\
  Effective HDM mass				& $m_\mathrm{s}^\mathrm{eff}/\mathrm{eV}$	& $[0,2]$			\\  \hline \hline
  \end{tabular}
  }
  \end{center}
\end{table*}

The $\Lambda$MDM model contains two additional base parameters to describe the HDM component: the effective number of extra neutrino species $\Delta \Neff$ and the effective sterile neutrino mass $m_\mathrm{s}^\mathrm{eff} = (94.1 \, \omega_\mathrm{s})$ eV with $\omega_\mathrm{s}= \omega_\text{hdm} =\Omega_\text{hdm} h^2$.
The former parameterises any contribution to the radiation energy density at photon decoupling by splitting the radiation density  into a sum  $\rho_\text{rad}=  \left( 1 + \Neff \frac{7}{8} \left(\frac{T_\nu}{T}\right)^4 \right) \rho_\gamma$ of the energy density in photons $\rho_\gamma$ and the energy density in SM neutrinos with the well-understood temperature ratio $T_\nu/T= (4/11)^{1/3}$ and $\Neff^\text{SM}=3.046$ such that any departure from the standard scenario shows up as $ \Delta \Neff \equiv \Neff - \Neff^\text{SM} \geq 0$.
In the considered case of a thermally distributed sterile neutrino, the effective sterile neutrino mass is related to its physical mass $m_\mathrm{s}$ via
\begin{equation}
 \label{mnuseff}
m_\mathrm{s}^\mathrm{eff} = (T_\mathrm{s}/T_\nu)^3 m_\mathrm{s} = (\Delta \Neff)^{3/4} m_\mathrm{s}\,.
\end{equation}
Note that both additional parameters, $\Delta \Neff$ and $m_\mathrm{s}^\mathrm{eff}$, can be mimicked by hot thermal relics of any nature and the considered case is equivalent for cosmological observables to a species distributed proportionally to active neutrinos.  
In the absence of further interactions, the observational effects rely on gravity, which is sensitive to the energy content only, and thus indifferent to the precise nature of the particles, or the details of their production, be it through oscillations from standard neutrinos or from thermalisation in a mirror sector~\cite{Abazajian:2012ys,Higaki:2013vuv}.  
We would like to emphasise that qualitatively different origins for the HDM like late cosmological particle decay~\cite{Hasenkamp:2012ii} also mimic a sterile neutrino species.

\subsection{Data}
We consider the following data sets:
\begin{itemize}
\item{{\bf CMB}: CMB TT angular power spectrum data from Planck \cite{Planck:2013kta}, combined with large-scale EE- and TE-polarisation power spectra from the 9-year WMAP data release~\cite{Bennett:2012zja}.}
\item{{\bf HST}: The measurement of the Hubble parameter using nearby type Ia supernova calibrated with observations of Cepheids by the Hubble Space Telescope, $(H_0 =  73.8 \pm 2.4)~\mathrm{km~s}^{-1}\mathrm{Mpc}^{-1}$~\cite{Riess:2011yx}.}
\item{{\bf C}: Cluster number counts from the Planck Sunyaev-Zeldovich catalog, approximately constraining the parameter combination $\sigma_8 \left(\Omega_\mathrm{m}/0.27\right)^{0.3} = 0.782 \pm 0.010$~\cite{Ade:2013lmv}.}
\item{{\bf BAO}: Measurements of the BAO scale by the 6dFGRS~\cite{Beutler:2011hx},
SDSS-II~\cite{Padmanabhan:2012hf}, and BOSS~\cite{Anderson:2012sa} surveys.}
\item{{\bf WL}: The 6-bin tomography angular galaxy shear power spectra from the CFHTLens survey, approximated as a constraint on $\sigma_8 \left(\Omega_\mathrm{m}/0.27\right)^{0.46} = 0.774 \pm 0.04$~\cite{Heymans:2013fya}.}
\end{itemize}
Note that the constraints on $\sigma_8 \left(\Omega_\mathrm{m}/0.27\right)^{x}$ from both cluster counts and galaxy shear were derived assuming vanilla cosmology.  We emphasise that, strictly speaking, an application of these constraints to non-vanilla models is incorrect, and one should, in principle, fit the unprocessed observables, i.e., cluster counts or shear power spectra, instead.  In the sterile model, the uncertainties will almost certainly be larger, so our results for the sterile model involving these data sets should be seen as indicative.  Nonetheless, assuming the shift of the posterior mean of $\sigma_8 \left(\Omega_\mathrm{m}/0.27\right)^{x}$ is small compared to the uncertainty when going from vanilla to sterile, our conclusions regarding consistency of data combinations can be considered conservative.

While most of these data sets are subject to potential bias from insufficiently understood systematics, we will entertain the idea that the constraints listed above are in fact unbiased measurements with properly characterised uncertainties.

Tension, in addition to the outlined inconsistencies, has been noted with supernova compilations~\cite{Conley:2011ku}, which, independent of the CMB, found best-fit values $\Omega_\text{m}= 0.223$-$0.227$ in the vanilla model. This is smaller than the values derived from CMB data.
Anyway, we find that supernovae cannot add substantial information, because they are overwhelmed in the combination by the other data.
Combined with the CMB data we found a marginally better fit in the extended cosmological model only.
Consequently, we do not consider them in our full data combination.
Since high multipole CMB data from SPT~\cite{Hou:2012xq} and ACT~\cite{Sievers:2013ica} appears consistent with Planck CMB data, we do not expect the inclusion of these data to affect our conclusions.
Recent fits of the full galaxy power spectrum in combination with Planck data have yielded tight constraints on the sum of neutrino masses \cite{Riemer-Sorensen:2013jsa}, but are also subject to systematic uncertainties regarding the modelling of non-linear corrections and scale-dependent galaxy bias.

We infer posterior probability distributions using the \texttt{CosmoMC} Markov-chain Monte Carlo package~\cite{Lewis:2002ah}.  For finding best-fits, we use the \texttt{BOBYQA} maximisation routine provided in \texttt{CosmoMC}.

\section{Results \label{sec:results}}
\begin{table*}[t]
  \caption{Best fit effective $\chi^2$ for various combinations of data sets in the vanilla (v) and sterile (s) models: total and individual contributions. \label{tab:bestfits}}
\begin{center}
{\footnotesize
  \hspace*{0.0cm}\begin{tabular}
  {c|ccccccc} \hline \hline
  Data & $-2 \ln \mathcal{L}_\mathrm{max}^\mathrm{tot}$ & $-2 \ln \mathcal{L}_\mathrm{max}^\mathrm{CMB}$ & $-2 \ln \mathcal{L}_\mathrm{max}^\mathrm{HST}$ & $-2 \ln \mathcal{L}_\mathrm{max}^\mathrm{C}$ & $-2 \ln \mathcal{L}_\mathrm{max}^\mathrm{BAO}$ & $-2 \ln \mathcal{L}_\mathrm{max}^\mathrm{WL}$ & Model \\       \hline
  \multirow{2}{*}{CMB} 		& 9802.5 & 9802.5 & --- & --- & --- & --- & v \\
   				  		& 9802.3 & 9802.3 & --- & --- & --- & --- & s \\ \hline
  \multirow{2}{*}{CMB+HST}	& 9808.4 & 9803.6 & 4.8 & --- & --- & --- & v \\
   				  		& 9803.2 & 9802.4 & 0.8 & --- & --- & --- & s \\ \hline	
  \multirow{2}{*}{CMB+C}		& 9818.1 & 9815.3 & --- & 2.8 & --- & --- & v \\
   				  		& 9806.5 & 9806.3 & --- & 0.1 & --- & --- & s \\ \hline
  \multirow{2}{*}{CMB+BAO}	& 9804.1 & 9802.7 & --- & --- & 1.4 & --- & v \\
   				  		& 9804.0 & 9802.3 & --- & --- & 1.8 & --- & s \\ \hline	
  \multirow{2}{*}{CMB+WL}	& 9808.5 & 9804.2 & --- & --- & --- & 4.3 & v \\
   				  		& 9806.4 & 9804.5 & --- & --- & --- & 1.9 & s \\ \hline	
  \multirow{2}{*}{CMB+all}		& 9825.2 & 9811.3 & 2.0 & 4.6 & 6.7 & 0.6 & v \\
  			 	  		& 9812.0 & 9807.4 & 2.2 & 0.2 & 1.7 & 0.5 & s \\ \hline \hline
  \end{tabular}
  }
  \end{center}
\end{table*}
Our first key result is summarised in Tab.~\ref{tab:bestfits} that provides for comparison the best fit effective $\chi^2$ values for the data combinations under consideration in the vanilla and sterile model.
We can see that the CMB data are inconsistent with HST, the cluster counts and the galaxy shear constraint as any of them combined with the CMB data leads to a $\Delta \chi^2_\mathrm{eff} > 2$ per additional degree of freedom. The situation is contrary for the sterile model. The $\Delta \chi^2$s are considerably smaller for any of these data combinations.
The BAO data, adding three degrees of freedom, can be combined consistently with the CMB in both models. Altogether, we find that all the data can be combined consistently within $\Lambda$MDM.
The last line of Tab.~\ref{tab:bestfits} shows how the extended model leads for the full data combination to a fit with an increase in quality by $\Delta \chi^2 = -13.2$ compared to the vanilla model.

If we combine the CMB data with all other data sets (CMB+all), we obtain our second key result, the following means and standard deviations:
\begin{align}
\label{nusparameter}
\Delta N_\mathrm{eff} &= 0.61 \pm 0.30 \nonumber \\
m_\mathrm{s}^\mathrm{eff} &=  (0.41 \pm 0.13)~\mathrm{eV}.
\end{align}
We caution that the statistical significance of these results may be somewhat overestimated, due the approximation in modelling the cluster and lensing likelihoods.
Exploiting~\eqref{mnuseff}, the physical neutrino mass corresponding to our mean values is $m_\mathrm{s} \simeq 0.59~\text{eV}$. 
The $\Delta\Neff=1$ of a fully thermalised neutrino species is well within the $2$-$\sigma$ range of our result. 
The $\Delta m^2 \sim 0.5~\text{eV}^2$ motivated by the accelerator anomaly corresponds, assuming $\Delta\Neff=1$, to a $m_\mathrm{s}^\mathrm{eff} \simeq 0.71~\text{eV}$, which is in tension with~\eqref{nusparameter} at slightly more than $2$-$\sigma$.
The mean of $m_\mathrm{s}^\mathrm{eff}$ is more than $4.5$-$\sigma$ below the expectation from a fully-thermalised sterile neutrino with $m_\mathrm{s}=1~\text{eV}$.

\begin{figure}[th]
\center
\includegraphics[height=\textwidth,angle=270]{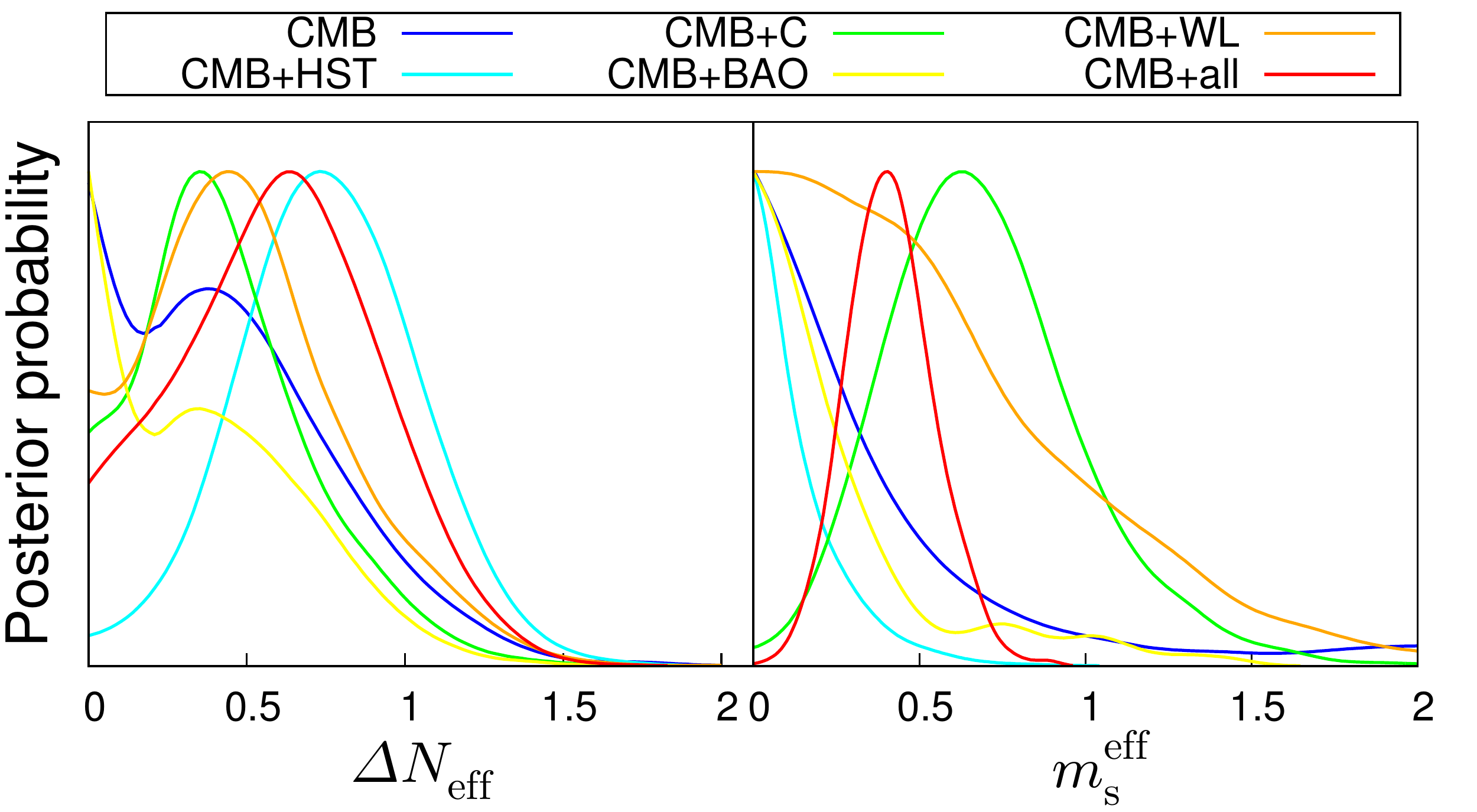}
\caption{Posterior probabilities (normalised to their respective maximum values) for the extra parameters of the sterile model.  \label{fig:nu1d}}
\end{figure}
In the left panel of Fig.~\ref{fig:nu1d} we can see that the evidence for $\Delta\Neff >0$ is mainly driven by the large HST value of $H_0$ as the posterior probability distribution peaks most far away from zero in the CMB+HST combination, in accordance with our expectation based on the positive correlation between the two parameters, as explained in Refs.~\cite{Bashinsky:2003tk,Hou:2011ec,Julienbook}. Galaxy shear and cluster constraints seem to slightly prefer a  non-zero value, too, even though significantly smaller than one.
The CMB only posterior very slightly prefers $\Delta\Neff>0$ which leads to a suppression of the power spectrum at higher multipoles due to stronger Silk damping~\cite{Hou:2011ec}.\footnote{This is in accordance with Planck's CMB-only best-fit for liberated $\Neff$ which is $\sim 1$-$\sigma$ above the SM expectation~\cite{Ade:2013zuv}.
Earlier analyses of high multipole data from the South Pole Telescope~\cite{Hou:2012xq} combined with WMAP7~\cite{Larson:2010gs} have indicated an increase at the same level of significance.}
The constraint arising from the CMB+BAO combination just does not exclude increases in $\Neff$ as small as in~\eqref{nusparameter}.

The main driver of the evidence for $m_\mathrm{s}^\mathrm{eff} > 0$ can be identified as the galaxy cluster counts in the right panel of Fig.~\ref{fig:nu1d}. The sterile neutrino does not cluster below its free-streaming scale, suppressing structure formation on small scales and thus yielding smaller values of $\sigma_8$, closer to the cluster count mean.  At the same time, lower values of $\omega_\mathrm{m}$ conspire to further suppress $\sigma_8$, and combined with smaller $\Omega_\mathrm{m}$, $\sigma_8 \left(\Omega_\mathrm{m}/0.27\right)^{0.3}$ can be brought down to the region favoured by the cluster data.  The lensing data, on the other hand, are much less constraining than the cluster data, and do not manage to dislodge the matter density from its CMB-preferred value.  Here, the improvement in the fit to $\sigma_8 \left(\Omega_\mathrm{m}/0.27\right)^{0.46}$ is achieved solely by allowing a larger $m_\mathrm{s}^\mathrm{eff}$. The other single combinations and the CMB-only data lead to relatively tight upper bounds on the neutrino mass. This is consistent with the results of Ref.~\cite{Verde:2013cqa}, who consider neither the Planck cluster data nor the CFHTLenS weak lensing data, and do not find any preference for a non-zero neutrino mass.

\begin{figure}[th]
 
\center
\includegraphics[width=\textwidth,angle=0]{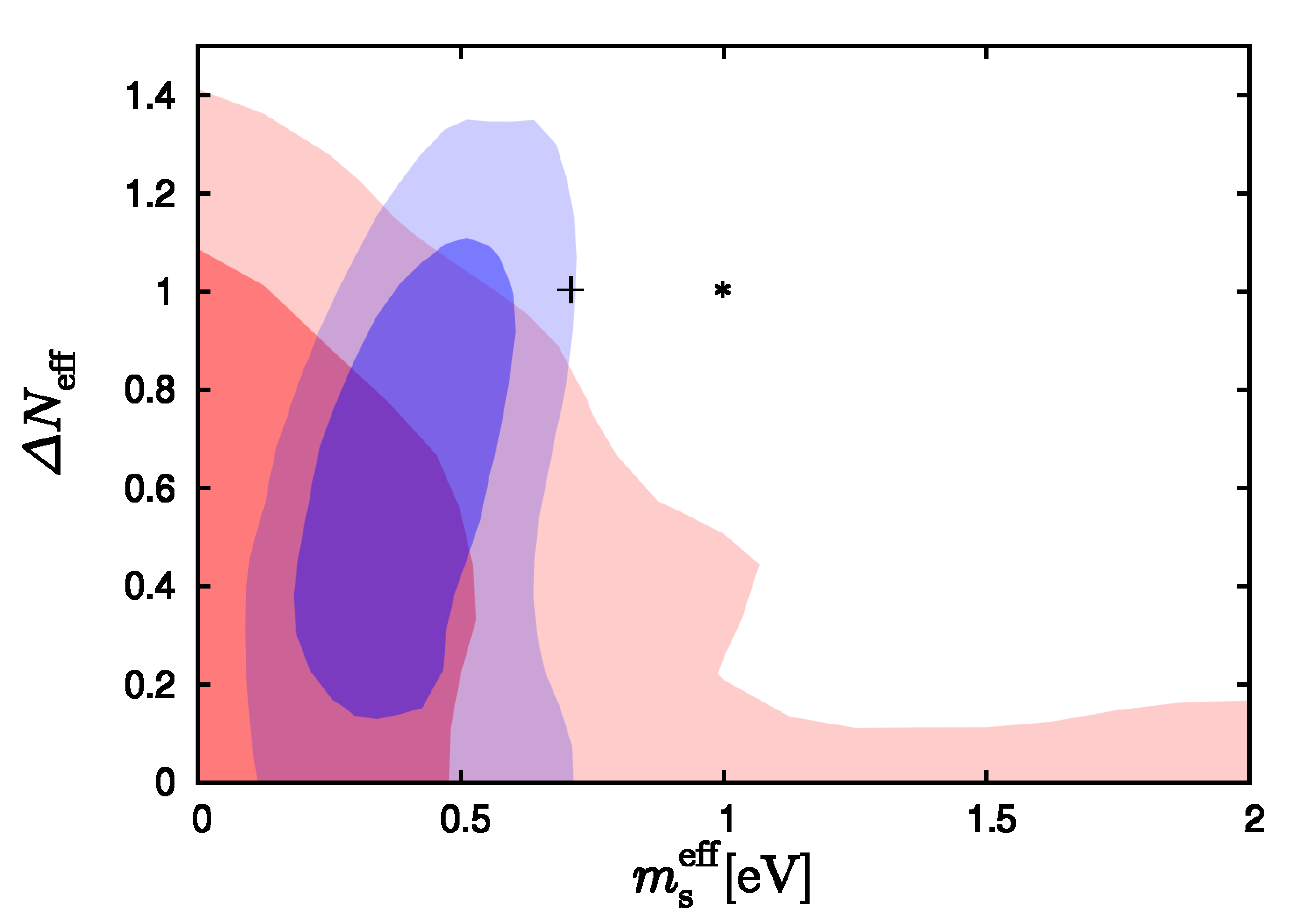}
\caption{Joint 68\%- and 95\%-credible contours of the marginalised posterior for the extra parameters of the sterile model.  Red contours correspond to CMB data only,
while the blue contours represent the full data combination. The vanilla model is located at the origin. Fully thermalised neutrinos correspond to $\Delta\Neff=1$ by construction.
Parameter values corresponding to a low mass sterile neutrino as motivated by the reactor and gallium anomalies are marked with an asterisk (*). A small region around the point marked by a cross (+) is motivated by the accelerator anomaly.
\label{fig:nu2d}}
\end{figure}
In Fig.~\ref{fig:nu2d} we compare the credible contours of the marginalised posterior probabilities for $\Delta\Neff$ and $m_\mathrm{s}^\mathrm{eff}$ from CMB data only with the full data combination. The corresponding one-dimensional intervals for the full data combination are given in~\eqref{nusparameter}.
We see that the CMB data yields, first of all, upper bounds on both parameters, while in the full data combination the best-fit point moves considerably away from zero. 
The CMB data alone allows for the tail towards large $m_\mathrm{s}^\mathrm{eff}$ at very low $\Delta\Neff$.  This region corresponds to relatively large physical $m_\mathrm{s}$, which becomes, for the CMB, indistinguishable from the CDM component.  This tail is also responsible for the multimodal nature of some of the 1-dimensional posteriors in the left panel of Fig.~\ref{fig:nu1d}.

The overlap between both contours is substantial. This is simply due to the fact that the contour of the full data combination lies at small $\Delta\Neff \lesssim 1$ and $m_\mathrm{s}^\mathrm{eff}$ so small that the CMB loses major parts of its sensitivity to them.
Most interestingly, we can infer from Fig.~\ref{fig:nu2d} that the vanilla model, located at the origin, is rejected at $3$-$\sigma$ if all data are combined. 

\begin{figure}[th]
 
\center
\includegraphics[width=\textwidth,angle=0]{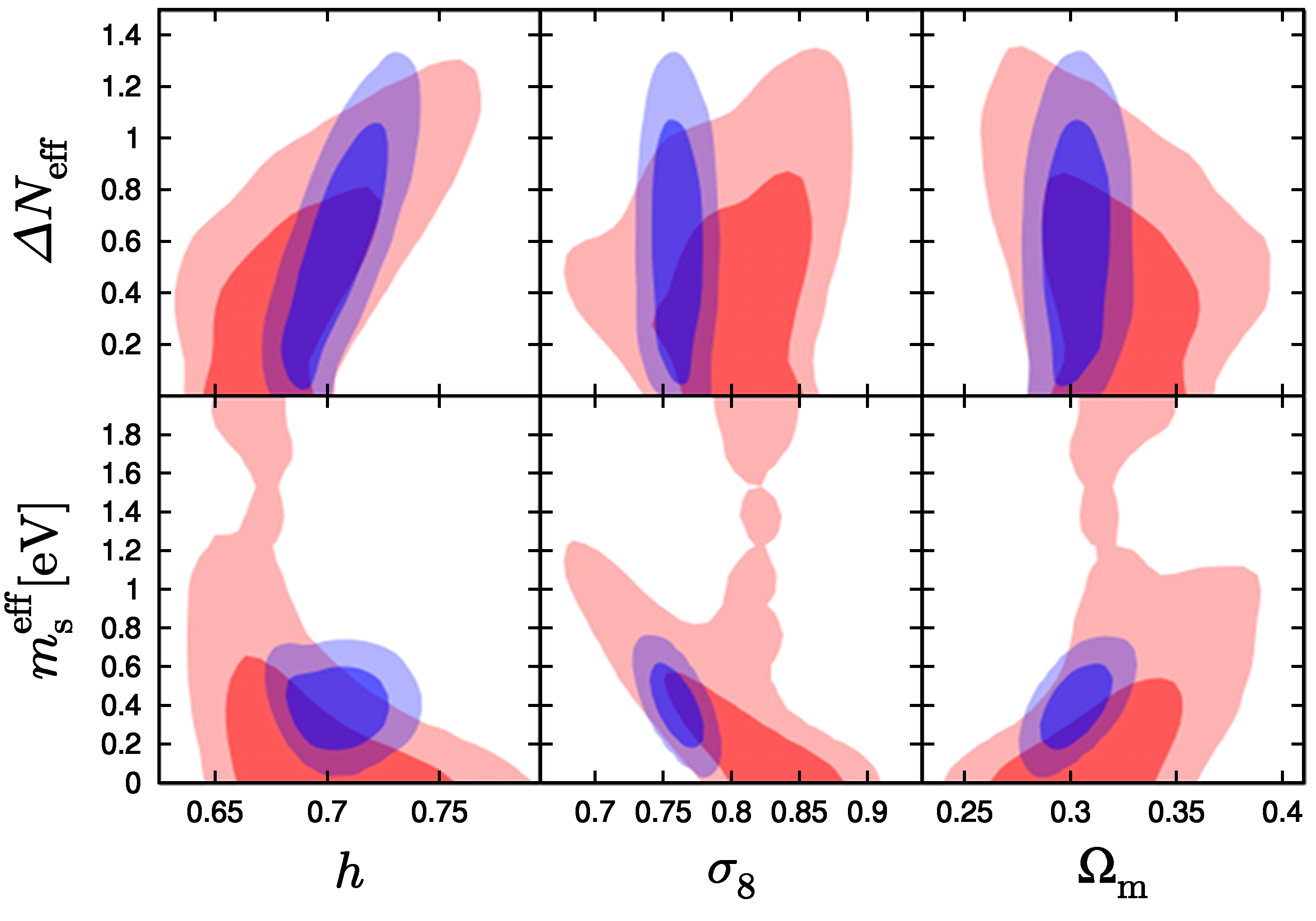}
\caption{Joint 68\%- and 95\%-credible contours of the marginalised posterior, illustrating the correlations between the extra parameters of the sterile model and the parameters $\sigma_8$, $H_0$ and $\Omega_\mathrm{m}$.  Red contours correspond to CMB data only, blue contours represent the full data combination. 
\label{fig:degeneracies}}
\end{figure}

Even though the non-CMB data sets do not directly constrain $\Delta N_\mathrm{eff}$ and $m_\mathrm{s}^\mathrm{eff}$, they nonetheless do so indirectly, by breaking parameter degeneracies the CMB data are subject to.  We illustrate this effect in Fig.~\ref{fig:degeneracies}.  The physical origin of the degeneracy directions introduced by allowing the radiation density or the sum of neutrino masses to vary can be understood by looking at parameter combinations which leave the main observed features of the angular power spectrum unchanged, and we refer the reader to Ref.~\cite{Julienbook} or the appendix of Ref.~\cite{Archidiacono:2013cha} for an in-depth discussion of this issue.  Note in particular the fact that the Hubble parameter is positively correlated with $\Delta N_\mathrm{eff}$, but anti-correlated with $m_\mathrm{s}^\mathrm{eff}$.  Thus, fitting HST and cluster data at the same time, which leads to an increase $\Delta N_\mathrm{eff}$ and $m_\mathrm{s}^\mathrm{eff}$, will introduce a small degree of tension within the CMB data (resulting in a deterioration of $\Delta \chi^2_\mathrm{eff} \approx 5$ in the best-fit to the CMB data when all data sets are combined compared to fitting CMB data alone).

To resolve the reactor and gallium anomalies mixing angles $\sin^2{2 \Theta} \gtrsim 0.1$~\cite{Abazajian:2012ys} are preferred. Regarding the current experimental results, mixing angles preferred by the accelerator anomaly reside in a small region around $\sin^2{2 \Theta} \sim 5 \times 10^{-3}$~\cite{Antonello:2012pq}. With these parameter values, one generically expects full thermalisation of the sterile neutrinos in the early universe, resulting in a contribution $\Delta N_\mathrm{eff} = 1$. 
We see from~\eqref{mnuseff} that in this case $m_\mathrm{s}^\mathrm{eff} = m_\mathrm{s}$. 
Concerning the motivation for a reactor or gallium anomaly neutrino, we can see that the corresponding point marked with an asterisk at $\Delta\Neff=1$ and $m_\mathrm{s}^\mathrm{eff} =1\text{ eV}$ is rejected even more strongly, i.e., with an even higher statistical significance than the vanilla model.
A sterile neutrino as motivated by the accelerator anomaly, $\Delta\Neff=1$ and $m_\mathrm{s}^\mathrm{eff} \simeq 0.71\text{ eV}$, appears in tension at the $2$-$\sigma$ level with our combined data set.
Note that the production of sterile neutrinos in the early universe can be suppressed by, e.g., an initial lepton asymmetry~\cite{Hannestad:2012ky,Mirizzi:2013kva}, permitting arbitrary values $0 \leq \Delta N_\mathrm{eff} \leq  1$.

\begin{table*}[t]
  \caption{Means and standard deviations of selected parameter posterior distributions for various combinations of data sets in the vanilla (v) and sterile (s) models. \label{tab:constraints}}
\begin{center}
{\footnotesize
  \hspace*{0.0cm}\begin{tabular}
  {c|cccccc} \hline \hline
  Data & $\Omega_\mathrm{m}$ &  $n_\mathrm{s}$ & $H_0$ & $\sigma_8 (\Omega_\mathrm{m}/0.27)^{0.3}$ & $\sigma_8 (\Omega_\mathrm{m}/0.27)^{0.46}$ & Model \\       \hline
  \multirow{2}{*}{CMB} 		& $0.316 \pm 0.017$ & $0.960 \pm 0.007$ & $67.3 \pm 1.2$ & $0.869 \pm 0.023$ & $0.891 \pm 0.031$ & v \\
   				  		& $0.321 \pm 0.027$ & $0.973 \pm 0.015$ & $68.8 \pm 2.8$ & $0.845 \pm 0.034$ & $0.868 \pm 0.036$ & s \\ \hline
  \multirow{2}{*}{CMB+HST}	& $0.299 \pm 0.014$ & $0.967 \pm 0.007$ & $68.5 \pm 1.1$ & $0.848 \pm 0.021$ & $0.861 \pm 0.027$ & v \\
   				  		& $0.297 \pm 0.018$ & $0.989 \pm 0.011$ & $72.0 \pm 2.1$ & $0.853 \pm 0.030$ & $0.866 \pm 0.031$ & s \\ \hline	
  \multirow{2}{*}{CMB+C}		& $0.270 \pm 0.008$ & $0.975 \pm 0.006$ & $70.6 \pm 0.8$ & $0.796 \pm 0.009$ & $0.797 \pm 0.012$ & v \\
   				  		& $0.305 \pm 0.022$ & $0.969 \pm 0.014$ & $67.5 \pm 2.2$ & $0.786 \pm 0.010$ & $0.813 \pm 0.014$ & s \\ \hline
  \multirow{2}{*}{CMB+BAO}	& $0.309 \pm 0.010$ & $0.963 \pm 0.006$ & $67.8 \pm 0.8$ & $0.860 \pm 0.022$ & $0.879 \pm 0.022$ & v \\
   				  		& $0.308 \pm 0.012$ & $0.976 \pm 0.011$ & $69.5 \pm 1.8$ & $0.846 \pm 0.030$ & $0.864 \pm 0.032$ & s \\ \hline	
  \multirow{2}{*}{CMB+WL}	& $0.295 \pm 0.012$ & $0.968 \pm 0.007$ & $68.8 \pm 1.0$ & $0.837 \pm 0.018$ & $0.849 \pm 0.023$ & v \\
   				  		& $0.322 \pm 0.033$ & $0.974 \pm 0.016$ & $68.8 \pm 3.0$ & $0.799 \pm 0.035$ & $0.821 \pm 0.030$ & s \\ \hline	
  \multirow{2}{*}{CMB+all}		& $0.279 \pm 0.007$ & $0.973 \pm 0.005$ & $70.0 \pm 0.6$ & $0.800 \pm 0.008$ & $0.804 \pm 0.010$& v \\
  			 	  		& $0.303 \pm 0.011$ & $0.986 \pm 0.011$ & $70.6 \pm 1.5$ & $0.786 \pm 0.010$ & $0.801 \pm 0.011$ & s \\ \hline \hline
  \end{tabular}
  }
  \end{center}
\end{table*}
In the first line of Tab.~\ref{tab:constraints} we see that without any additional data the mean values of $H_0$ and the parameter combinations $\sigma_8 \left(\Omega_\mathrm{m}/0.27\right)^{x}$ move on the $1$-$\sigma$ level in the extended model towards values as found in the additional data. Moreover, we see that the errors on all parameters grow substantially. In the sterile model the CMB error dominates in every observable over the error in the corresponding local observation.
If we compare the parameter values determined from the CMB alone in the vanilla model (uppermost line) with those determined from the full data combination in the sterile model (lowermost line), the drastic shift towards the local measurements becomes obvious. Indeed, we find a perfectly reasonable fit for the two $\sigma_8 \left(\Omega_\mathrm{m}/0.27\right)^{x}$ constraints. Interestingly, the mean value of $H_0$ is still smaller than the HST value, but since the corresponding error is relatively larger than in the vanilla model, there is no strong tension in the sterile model.
The second column of Tab.~\ref{tab:constraints} demonstrates how the addition of the non-CMB data sets pushes the scalar spectral index $n_\mathrm{s}$ to larger values in order to uphold the fit to the CMB power spectrum. 
We would like to remind the reader that, when comparing parameter posterior distributions of the vanilla with the sterile model,  Tab.~\ref{tab:constraints} does not show the increase and differences in the quality of the fit. These are provided in Tab.~\ref{tab:bestfits}.

As a side note, it is interesting that after combining all the data $\Omega_\text{m}$ is -- against na\"ive intuition -- reduced by $\sim3$\%, in the sterile model even though the model comes with an additional matter component. This might ameliorate the tension between the mentioned supernova compilations and the CMB within $\Lambda$MDM if the additional data are included.

\section{Conclusions \label{sec:conclusions}}
We have shown that the $\Lambda$MDM model, in contrast to the vanilla model, allows to combine the new high-precision CMB data with local $H_0$ measurements and determinations of $\sigma_8 \left(\Omega_\mathrm{m}/0.27\right)^{x}$ from both, cluster counts and galaxy shear, cf.~Tab.~\ref{tab:bestfits}.
In our full data combination $\Lambda$MDM provides a a considerably better fit, $\Delta \chi^2 = -13.2$, than the vanilla model.
Our results show statistical evidence for a HDM component described by a sterile neutrino species with cosmological parameters more than $2\sigma$ above the default zeros of the vanilla model, see~\eqref{nusparameter}.
In the case of $\Delta\Neff$ the preference is mainly driven by the large local $H_0$ value,
while cluster counts drive the preference for a non-zero mass.
Galaxy shear data prefers a small $\Delta\Neff>0$ as well as a non-zero mass. 
The combined two-dimensional posterior probability distribution, Fig.~\ref{fig:nu2d}, hints at new particle physics at the $3\sigma$-level, while sterile neutrinos as motivated by the reactor and gallium anomalies seem rejected at even higher significance than the vanilla model.
A sterile neutrino motivated by the accelerator anomaly appears in $2\sigma$-tension with cosmological data.

While our results strengthen the case for light sterile neutrinos as motivated by anomalies in neutrino experiments relative to conclusions from CMB data alone, the remaining tension (or rejection at high significance, respectively) of straightforward expectations in simple models might disfavour these scenarios and motivate more sophisticated attempts to reconcile cosmological and experimental data.
Any more, our results call for other candidates, origins and explanations for a hot dark matter component that contributes somewhat less than a fully thermalised relic to the radiation energy density at photon decoupling and becomes non-relativistic at the right time to provide the desired non-relativistic energy density today.

Next year's release of Planck's full mission data, including polarisation information, will shed more light on the issue.
It will also be interesting how other measurements providing similar constraints to those considered in our work, for example, from the power spectrum of the CMB lensing potential~\cite{Ade:2013tyw} relate to our findings as soon as their errors decrease with additional data.
Finally, with the information gathered by the next generation of large-volume galaxy surveys such as LSST or EUCLID, the sensitivity to the hot dark matter parameter space will greatly increase~\cite{Hamann:2012fe,Audren:2012vy}, allowing us to settle the question once and for all.

The reader should be aware that the data sets driving the evidence for the $\Lambda$MDM model are subject to numerous astrophysical systematics, which, if not modelled correctly, can lead to significant biases in parameter estimates.  However, if we assume that the systematic effects have been treated properly in these data, we find it remarkable that the extension of the Standard Model of particle physics by only one particle, 
allows to combine three additional observations with the cosmic microwave background, bringing them in concordance.
At second glance, the Universe might turn out less boring than initially thought, and the vanilla $\Lambda$CDM model may not have been the last word.

\paragraph*{Note added:}
During the finalisation of this work, Ref.~\cite{Wyman:2013lza}, which explores a very similar scenario, appeared on the arXiv.  Even though a direct, quantitative comparison is not possible, their results and conclusions agree qualitatively with ours.
Due to having to undergo a Planck collaboration internal approval process, the appearance of this paper on the arXiv has been delayed by 10 days.

\section*{Acknowledgements}
We thank Steen Hannestad and Antony Lewis for discussions and acknowledge the use of computing resources from the Danish Center for Scientific Computing (DCSC).
J.~Hasenkamp was supported by the German
Research Foundation (DFG) via the Junior Research Group ''SUSY Phenomenology`` within
the Collaborative Research Centre 676 ''Particles, Strings and the Early Universe`` and would
like to thank the German Academy of Science for support through the Leopoldina Fellowship
Programme grant LPDS 2012-14.

We acknowledge the use of Planck data. The development of Planck has been supported by: ESA; CNES and CNRS/INSU-IN2P3-INP (France); ASI, CNR, and INAF (Italy); NASA and DoE (USA); STFC and UKSA (UK); CSIC, MICINN and JA (Spain); Tekes, AoF and CSC (Finland); DLR and MPG (Germany); CSA (Canada); DTU Space (Denmark); SER/SSO (Switzerland); RCN (Norway); SFI (Ireland); FCT/MCTES (Portugal); and The development of Planck has been supported by: ESA; CNES and CNRS/INSU-IN2P3-INP (France); ASI, CNR, and INAF (Italy); NASA and DoE (USA); STFC and UKSA (UK); CSIC, MICINN and JA (Spain); Tekes, AoF and CSC (Finland); DLR and MPG (Germany); CSA (Canada); DTU Space (Denmark); SER/SSO (Switzerland); RCN (Norway); SFI (Ireland); FCT/MCTES (Portugal); and PRACE (EU).

A description of the Planck Collaboration and a list of its members, including the technical or scientific activities in which they have been involved, can be found at \linebreak \texttt{http://www.sciops.esa.int/index.php?project=planck\&page=Planck\_Collaboration}.


\end{document}